\documentclass{elsart}

\usepackage{graphicx}

\usepackage{amssymb}

\begin{document}

\begin{frontmatter}



\title{Ambiguity of gamma-ray tracking of `two-interaction' events}


\author{N.J.~Hammond},
\author{T.~Duguet},
\ead{duguet@theory.phy.anl.gov}
\author{C.J.~Lister}
\address{Physics Division, Argonne National Laboratory, Argonne, Illinois, 60439, USA}
\begin{abstract}
Tracking of gamma-ray interactions in germanium detectors can allow
reconstruction of the photon paths, and is useful for many
applications. Scrutiny of the kinematics and geometry of gamma rays
which are Compton scattered only once prior to full absorption reveals that
there are cases where even perfect spatial and energy resolution
cannot resolve the true interaction sequence and consequently gamma-ray
tracks cannot be reconstructed. The photon energy range where this
ambiguity exists is from $255$~keV to around $700$~keV. This is a
region of importance for nuclear structure research where two-point
interactions are probable.
\end{abstract}

\end{frontmatter}

\section{Introduction}
\label{Introduction}
Experimental nuclear structure research will
increasingly involve the use of radioactive ion beams to produce
nuclei far from stability. Due to the technical difficulties
associated with accelerating such beams, the anticipated intensities
are much lower than those currently achieved with stable-ion
beams. This will expose the efficiency shortcomings of the current
generation of gamma-ray spectrometers, such as
Gammasphere. Consequently, both in the U.S.~\cite{Greta} and in
Europe~\cite{Agata}, much attention has been focused on the development
of gamma-ray-tracking detectors which can provide much greater
efficiency and excellent doppler correction facility. The proposed
devices take the form of a spherical shell of highly-segmented germanium
detectors read out through digital electronics with sophisticated pulse-shape
analysis. Despite the difficulties in employing
this technology, a `proof of principle' has been published for
the proposed U.S. device Greta~\cite{Vetter} and progress continues to
be made~\cite{web}\\
Gamma-ray tracking is based upon the principle that when a photon
interacts within one segment of a germanium detector, the transient signals produced
in neighboring segments can be utilised to locate the interaction
point with great precision. Matching excellent spatial resolution with
the energy resolution afforded by germanium detectors, it is hoped
that the full interaction path, or track, can be deduced for incident gamma
rays. However, in discussion of this principle it has been
reported that given perfect spatial and energy resolution, the tracking
algorithm employed to decipher the events can identify the gamma-ray path
exactly~\cite{Vetter}. In this paper we show that there are exceptions
to this rule. We focus upon photons which are scattered only once prior
to full absorption to demonstrate that a general feature of Compton scattering leads to an
ambiguity in the tracking of particular events. As we perceive there
to be a general misconception that tracking ambiguities arise purely
from experimental uncertainties in energy or position, we hope that
this work will help to clarify the issue.\\

\section{Tracking gamma rays involved in two interactions}
\begin{figure}
\begin{center}
\includegraphics[width=0.5\textwidth,angle=0]{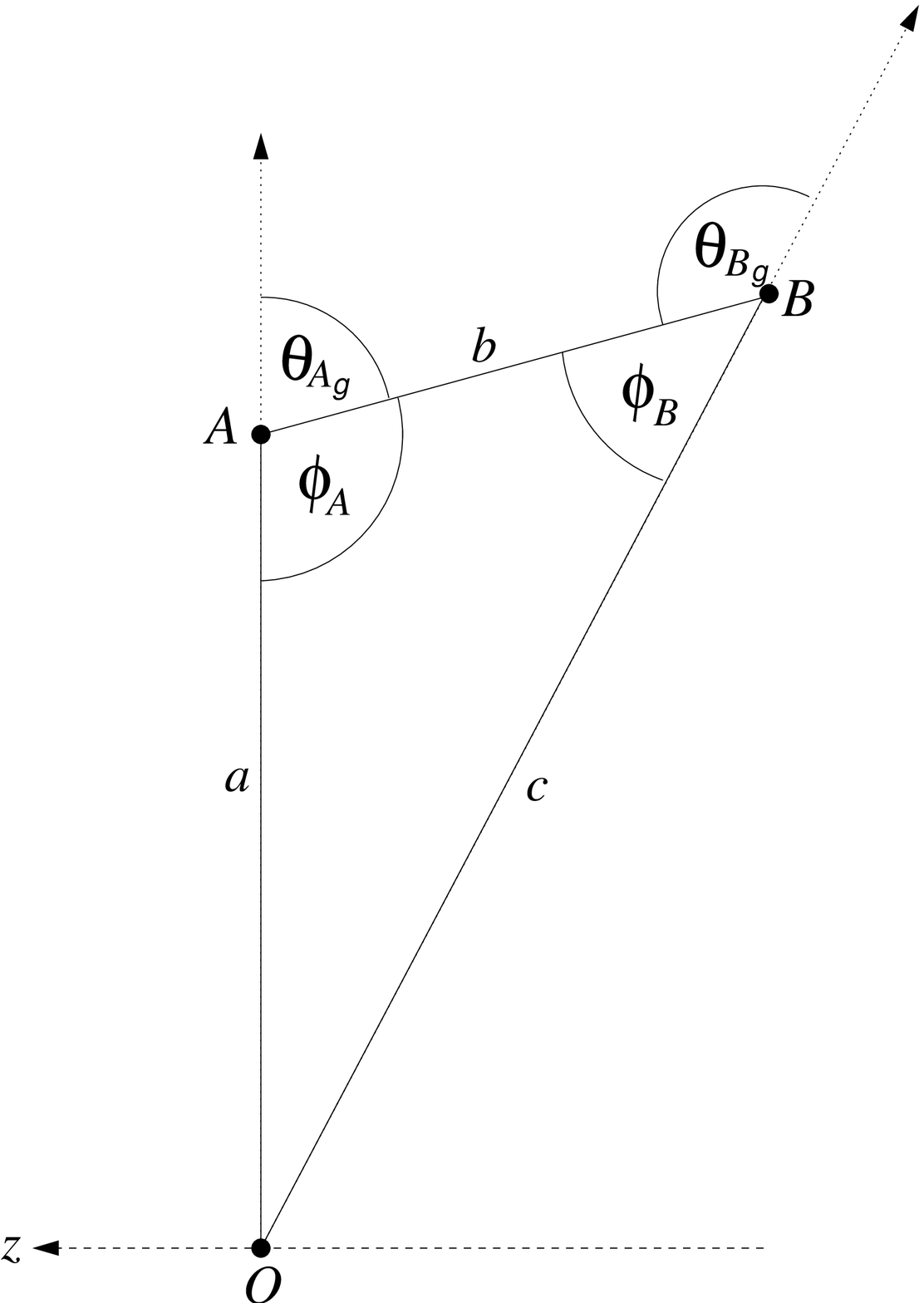}	
\end{center}
\caption{A figurative representation of a photon emitted from a
known origin O and subsequently interacting twice within matter at
points A and B. The two possible scattering angles, $\theta_{A_g}$ and
$\theta_{B_g}$, can be found from the geometry of the OAB triangle.}
\label{triangle}
\end{figure}

We restrict our discussion to gamma rays which undergo a single
Compton interaction prior to photoelectric absorption. For such
events, the process of identifying the gamma-ray track is a matter of
deducing the Compton scattering angle $\theta$. If the spatial origin
of the gamma ray is known, as it is approximately when studying decay from a radioactive source
or a target bombarded with ions, then this problem is further reduced
to a determination of which interaction occurred first. As the
intrinsic timing of germanium detectors is inadequate for this task,
precedence must be evaluated based upon a consideration of the
measured energies and the Compton scattering formula~\cite{Compton},
\begin{equation}
\frac{1}{E'_\gamma}-\frac{1}{E_\gamma}=\frac{1}{m_0c^2}(1-\cos\theta),   
\label{Compton}
\end{equation}
where $E'_\gamma$ is the incident photon energy, $E_\gamma$ is the
energy of the scattered photon and $\theta$ is the angle through which
the scattering occurred. 
To illustrate, we analyse an event where both interaction points and the
associated energies are known exactly. Labelling the interaction points $A$ and
$B$, and the partial energies deposited at each point $E_A$ and $E_B$
respectively, the incident energy is determined by
\begin{equation}
E_\gamma=E_A+E_B.
\end{equation}
However, as it is not known whether $A$ or $B$ is the point at which
the scattering occurred,  $E'_\gamma$ could be either $E_A$ or $E_B$. Rewriting
Eq.(\ref{Compton}) in terms of the angle $\theta$, and substituting for
$E_\gamma$ and $E'_\gamma$, we obtain two possibilities,
\begin{equation}
\theta_{A_k}=\cos^{-1}\left[1-511\left(\frac{{E_A}}{{E_B}({E_A}+{E_B})}\right)\right],\\
\label{solution1}
\end{equation}
or
\begin{equation}
\theta_{B_k}=\cos^{-1}\left[1-511\left(\frac{{E_B}}{{E_A}({E_A}+{E_B})}\right)\right],\\
\label{solution2}
\end{equation}
where $E_A$ and $E_B$ are in units of keV and the subscript $k$
signifies that these angles correspond to purely kinematic
solutions. Below a certain energy threshold, which we will deduce shortly,
only one physical solution exists and the gamma-ray track is uniquely
determined from consideration of the energies alone. However, above
the threshold, both Eq.(\ref{solution1}) and Eq.(\ref{solution2}) may
give physical solutions and the kinematic approach is insufficient,
hence the need for `tracking'. The two known interaction points, in
addition to the origin, define a triangle with geometry as shown in
Fig.\ref{triangle}. Solving this triangle yields the two `geometric'
solutions $\theta_{A_g}$ and $\theta_{B_g}$. 

Therefore, if A is the point at which the incident photon was
scattered, the Compton scattering angle is given by,
\begin{equation}
\theta_{A_g}=180^\circ-\phi_A=\cos^{-1}\left[(c^2-a^2-b^2)/2ab\right],
\label{thetaa}
\end{equation} 
whereas the alternative solution is given by,
\begin{equation}
\theta_{B_g}=180^\circ-\phi_B=\cos^{-1}\left[(a^2-b^2-c^2)/2bc\right].
\label{thetab}
\end{equation} 
Clearly, though there are two kinematic solutions and two geometric solutions, the
true scattering angle $\theta$ should be a solution common to both
approaches. If the two remaining solutions are not alike, then they
can be rejected and the tracking is known exactly. However, if
$\theta_{A_g}=\theta_{A_k}$ and $\theta_{B_g}=\theta_{B_k}$, then both
solutions are equally valid and the gamma-ray track is ambiguous. To
test whether such cases exist, we assume that
$\theta=\theta_{A_k}=\theta_{A_g}$, and attempt to identify solutions for
which $\theta_{B_k}=\theta_{B_g}$. For convenience, we introduce the
shorthand terms, 
\begin{equation}
X=\left[1-511\left(\frac{E_A}{E_B(E_A+E_B)}\right)\right],
\end{equation}
and
\begin{equation}
Y=\left[1-511\left(\frac{E_B}{E_A(E_A+E_B)}\right)\right],
\end{equation}
such that $\theta_{A_k}=\cos^{-1}X$ and $\theta_{B_k}=\cos^{-1}Y$. Rearranging
Eq.(\ref{thetaa}) in terms of $c$, replacing $\theta_{A_g}$ with
$\theta_{A_k}=\theta$, and substituting for $c$ in Eq.(\ref{thetab}) yields,
\begin{equation}
\theta_{B_g}=\cos^{-1}\left[-(2b^2+2baX)/2b\sqrt{(b^2+a^2+2baX)}\,\right].
\end{equation} 
Equating $\theta_{B_k}$ with $\theta_{B_g}$ gives, after further simplification,
\begin{equation}
b+aX+Y\sqrt{b^2+a^2+2baX}=0,
\end{equation} 
which is quadratic in $b$. Of the two solutions for $b$, one corresponds to the situation where
$\theta_{B_g}=180^\circ-\theta_{B_k}$. The other, for which
$\theta_{B_g}=\theta_{B_k}$ is given by,
\begin{equation}
b=a\left[-\cos\theta-\frac{E_\gamma}{511}\left(1-\cos\theta-\left(\frac{511}{E_\gamma}\right)^2\right)\sqrt{\frac{1-\cos^2\theta}{2(1-\cos\theta)-(\frac{511}{E_\gamma})^2}}\right].
\label{final}
\end{equation}
In physical terms, a positive solution to  Eq.(\ref{final}) indicates
that for a photon of energy $E_\gamma$, which travels a distance $a$
and is scattered through an angle $\theta$ for a further distance $b$
prior to absorption, the gamma-ray track cannot be deduced from the
locations of the two interaction points and the energies deposited.
The limits for which such solutions exist are given by,
\begin{equation}
\frac{1}{\sqrt{2}}\frac{511}{\sqrt{1-\cos\theta}}\leq E_\gamma\leq \frac{1}{\sqrt{1+\cos\theta}}\frac{511}{\sqrt{1-\cos\theta}}.
\label{lower}
\end{equation}

\section{Discussion}
\begin{figure}
\begin{center}
\includegraphics[width=\textwidth,angle=0]{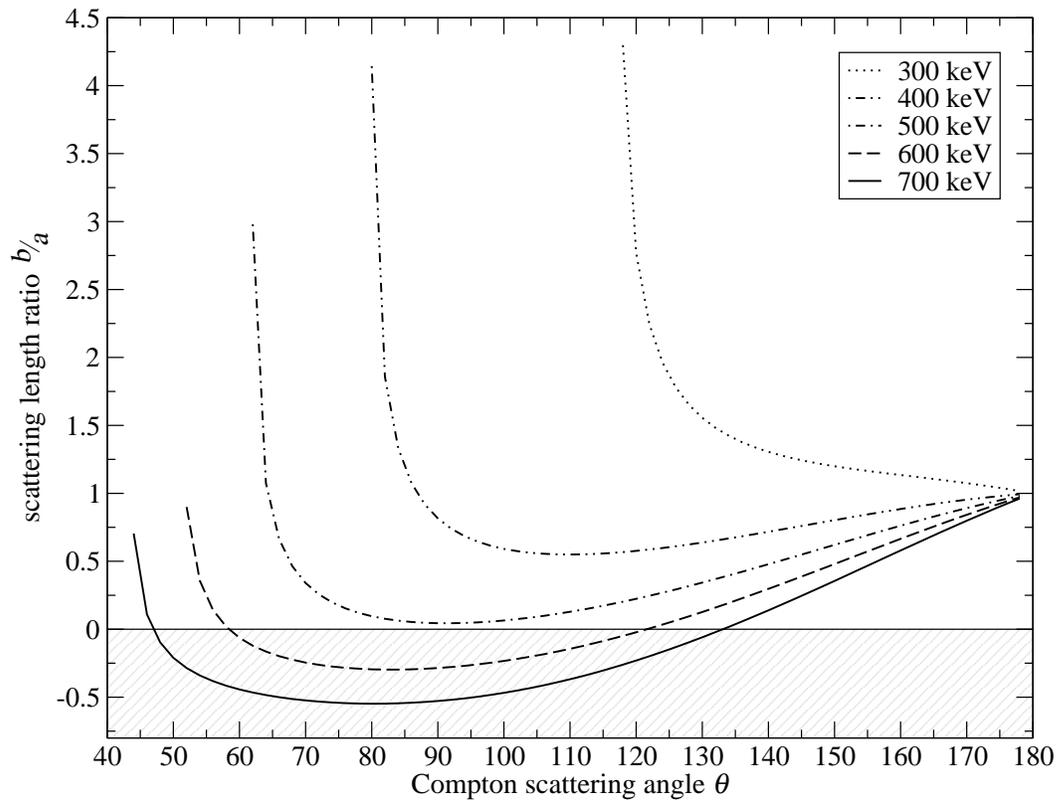}	
\end{center}
\caption{Solutions to Eq.(\ref{final}) given in terms of the ratio
$b/a$. Negative values, denoted by the shaded region,
correspond to non-physical solutions.}
\label{curves}
\end{figure}

In Fig.\ref{curves} the solutions to Eq.(\ref{final}) for $b/a$ are plotted against $\theta$ for
$E_\gamma=300\rightarrow700$ keV. From a study of this figure and
the conditions of Eq.(\ref{lower}), certain features are
evident. Clearly, as the energy increases beyond 511~keV, the range of
angles for which a tracking ambiguity may exist is reduced to larger
and smaller $\theta$ simultaneously. According to Eq.(\ref{lower}),
solutions exist up to
infinite energy as $\theta$ approaches $180^\circ$. 
At energies below 511~keV, the range of ambiguity is increasingly
restricted to higher $\theta$. This trend occurs rapidly; below
255~keV there are no angles for which an ambiguity exist.\\
Taking a single gamma-ray event in isolation, there are two primary physical
consequences of an ambiguity in the gamma-ray track. Firstly, the
angle of gamma-ray emission, needed to perform Doppler correction and
to measure angular distributions, cannot be deduced from two possibilities, which in Fig.~\ref{triangle}
correspond to the angles subtended by $a$ and $c$ to $z$. Secondly,
the linear polarisation of the gamma ray cannot be ascertained, as the
two possibilities $\theta_{A_g}$ and $\theta_{B_g}$ exist for the Compton
scattering angle. Except for cases where only the energy of the
incident gamma ray is of interest, the ambiguous events must
presumably be identified and then suppressed, resulting in a reduction
of efficiency.\\
Though there are too many factors affecting the
efficiency of a tracking array for the scope of this paper, we
can make some qualitative statements regarding this effect. Firstly,
if we make the likely assumption that the tracking array is
roughly spherical with a radius of $\sim10$~cm,
then solutions for which $b/a\geq1$ can reasonably be expected to
have a low probability, as they correspond to scattering lengths much
greater than the mean free path of a gamma ray in germanium. Secondly,
gamma rays with energies beyond 700~keV are unlikely to be involved in
only one interaction prior to scattering and will therefore generate
few ambiguous events. In addition, the probability of backward
scattering reduces dramatically with increasing energy and therefore
solutions for $\theta>90^\circ$ are improbable at energies above
around 700~keV. These features suggest that it is roughly the energy range
from 400~keV to 500~keV which is particularly susceptible to tracking
ambiguity. Notably, this is an energy region of prime importance to
nuclear structure research.\\
Our discussion has been based upon the principle of infinite spatial
and energy resolution. In practice neither of these conditions exist,
nor is the gamma-ray origin known with precision. This compounds
the problem, effectively broadening each curve of
Fig.\ref{curves}. Additionally, it is at small values of $a$ for which
the effect is most liable to become a problem. Unfortunately the cost
of building an array of segmented detectors is a strong motivation to
bring the detectors as close to the source as possible which increases
the probability of ambiguous events. Finally, the effect we have
described may lead to higher order problems when multiple gamma rays
are involved. Much work on the tracking algorithms has been
dedicated to the identification of separate gamma-ray tracks. This
added uncertainty may make such a task more difficult and needs to be
considered when appraising the overall response of a tracking device or the
overall efficiency of tracking algorithms. The analytical solutions
and boundary conditions derived herein should prove valuable in this task.

\section{Acknowledgments}

This work was supported by the US Department of Energy, Office of Nuclear Physics, under contract W-31-109-ENG38.

\newpage 

\end{document}